\documentclass[12pt,preprint]{aastex}
\usepackage{natbib}
\usepackage{graphicx}
\shorttitle{What if planet 9 has satellites?}
\begin{document}
\title{What if planet 9 has satellites?}
\author{Man Ho Chan}
\affil{Department of Science and Environmental Studies, The Education University of Hong Kong, Hong Kong, China}
\email{chanmh@eduhk.hk}

\begin{abstract}
In the past decade, numerical simulations started to reveal the possible existence of planet 9 in our solar system. The planet 9 scenario can provide an excellent explanation to the clustering in orbital elements for Kuiper Belt objects. However, no optical counterpart has been observed so far to verify the planet 9 scenario. Therefore, some recent studies suggest that planet 9 could be a dark object, such as a primordial black hole. In this article, we show that the probability of capturing large trans-Neptunian objects (TNOs) by planet 9 to form a satellite system in the scattered disk region (between the inner Oort cloud and Kuiper Belt) is large. By adopting a benchmark model of planet 9, we show that the tidal effect can heat up the satellites significantly, which can give sufficient thermal radio flux for observations, even if planet 9 is a dark object. This provides a new indirect way for examining the planet 9 hypothesis and revealing the basic properties of planet 9.  
\end{abstract}


\section{Introduction}
Currently, there are 8 planets officially identified in our solar system. Most of the newly discovered large astronomical objects outside Neptune are dwarf planets or large asteroids called trans-Neptunian objects (TNOs). In view of the TNOs, the new discovery of 2012 VP113 and some potential members of the inner Oort cloud has revealed a strange clustering in orbital elements \citep{Trujillo}. The perihelion distance have arguments of perihelia $\omega$ clustered approximately round zero \citep{Trujillo,Batygin}. Later analysis shows that the chance for this strange clustering due to random is just 0.0007\% \citep{Batygin}. Therefore, a dynamical mechanism involving a new planet located at more than 100 AU has been suggested \citep{Batygin2}. Many studies have constrained the mass and the orbital properties of the hypothesized planet 9 (P9) \citep{Batygin3,Sheppard,Gomes,Becker,Sheppard2}. Current benchmark models suggest that P9 has mass $M_9 \sim 5-10M_{\oplus}$, orbital semi-major axis $a_9 \sim 400-800$ AU and eccentricity $e_9 \sim 0.2-0.5$ \citep{Batygin2}. However, the in-situ formation of P9 is strongly disfavored so that P9 might be a captured planet from the free-floating objects nearby the solar system \citep{Batygin2,Kenyon}. A more detailed assessment of the probability of capture can be found in \citet{Li}.

Current benchmark models of P9 suggest that it has a temperature $\sim 40$ K and a radius $\sim 3-4R_{\oplus}$ \citep{Batygin2}. The possible location of P9 in the celestial sphere is also constrained \citep{Batygin2,Fienga,Socas}. Based on these properties, various observations, such as optical and microwave/infrared observations, have been deployed to observe the hypothesized P9 \citep{Meisner,Meisner2,Naess}. However, no electromagnetic wave signal has been detected for P9 \citep{Meisner,Meisner2,Linder}. Careful examinations based on previous optical surveys also do not reveal the existence of P9 \citep{Linder}. Therefore, these null results have made the P9 hypothesis more mysterious.

In view of these problems, some of the studies have suggested that P9 is a dark object (dark P9), such as a compact object made by dark matter \citep{Wang} or a primordial black hole (PBH) \citep{Scholtz}. In particular, the proposal of the PBH P9 has attracted many discussions because many studies beyond the standard models have already proposed the existence of PBHs with mass $\sim M_{\oplus}$. There are various mechanisms which can generate PBHs in early universe \citep{Carr}. However, the direct signals emitted by the PBH P9 (e.g. Hawking radiations) are too small to detect \citep{Arbey}. Even if we assume dark matter can distribute around the PBH P9, the resulting gamma-ray signals might be smaller than the current observation limits \citep{Scholtz}. Besides, a recent innovative proposal suggests that using a small laser-launched spacecraft with a velocity of order $0.001c$ can reach the PBH P9 to detect its gravitational field, though we need to wait for the measurement after roughly a decade \citep{Witten}.

Nevertheless, there are a lot of TNOs orbiting about the sun inside the scattered disk region ($\sim 100-1000$ AU), located between the inner Oort cloud and Kuiper Belt. These TNOs are also known as detached objects. Most of them are either scattered from the central solar system or Kuiper Belt region. In fact, we have already observed at least 47 large TNOs with orbital semi-major axis larger than 100 AU and size larger than 100 km. Therefore, it is possible that these large TNOs would be captured by P9 to become satellites of P9. Many dwarf planets such as Pluto and TNOs outside Neptune have satellite systems \citep{Brown,Grundy}. If these small objects can have satellites, it can be conceived that the more massive P9 might also have a number of satellites. In this article, we discuss some important observable features if P9 has captured satellites. For large satellites with small orbital semi-major axis, the tidal heating effect due to P9 would be important. It can be shown that these satellites would give an observable standard thermal radio spectrum. If P9 is a dark object, observing the satellites would be another kind of investigation to examine the P9 hypothesis in the near future. In the followings, we assume that P9 is a dark object and we follow the benchmark model of P9 with mass $M_9=5M_{\oplus}$, eccentricity $e_9=0.2$, orbital inclination $i=20^{\circ}$, and semi-major axis $a_9=450$ AU \citep{Batygin2}. We simply take the semi-major axis $a_9=450$ AU as the average distance to the dark P9 from the Earth.

\section{Capturing probability}
There are many large TNOs moving in the scattered disk region ($\sim 100-1000$ AU), such as 2018 AG37, 2018 VG18 and 2020 BE102. It is quite likely that some of the large TNOs (e.g. with size $D \sim 100$ km) could be captured by the dark P9. In fact, many of the Kulper Belt dwarf planets have at least one satellite. For example, the satellite of the dwarf planet Eris has radius $R \sim 700$ km and semi-major axis $a \sim 4 \times 10^4$ km \citep{Brown2}.

In general, when a TNO has a close encounter to a planet, energy will be lost in the capturing process due to the inverse of the gravitational slingshot mechanism \citep{Napier}. The maximum capturing distance between the dark P9 and any TNOs can be characterized by the impact parameter $b$ \citep{Napier}:
\begin{equation}
b \sim \frac{M_9}{M_{\odot}} \left(\frac{GM_{\odot}}{a_9}\right)^{3/2}v^{-3}a_9,
\end{equation}
where $v$ is the incoming relative velocity between the dark P9 and any TNOs. Here, $b$ can be regarded as the closest distance between the dark P9 and the TNOs for the capturing process. Therefore, the relative velocity between the dark P9 and the TNOs is given by
\begin{equation}
v \sim \sqrt{\frac{GM_{\odot}}{a_9}}-\sqrt{\frac{GM_{\odot}}{a_9 \pm b}} \cos \Delta i,
\end{equation}
where $\Delta i$ is the orbital inclination difference between the dark P9 and the TNOs. As $b \ll a_9$, the relative velocity is
\begin{equation}
v \sim \sqrt{\frac{GM_{\odot}}{a_9}}(1- \cos \Delta i).
\end{equation}
Putting Eq.~(3) into Eq.~(1), we get
\begin{equation}
b \sim a_9(1-\cos \Delta i)^{-3} \left(\frac{M_9}{M_{\odot}} \right).
\end{equation}
The benchmark orbital inclination of the dark P9 is $i=20^{\circ}$ \citep{Batygin2}. Based on the catalog compiled by the International Astronomical Union \footnote{The catalog compiled by the International Astronomical Union can be found in https://minorplanetcenter.net/iau/lists/TNOs.html}, the orbital inclinations of the TNOs (with semi-major axis $a>100$ AU) are quite close to $i=20^{\circ}$, except three with $i>100^{\circ}$. The average difference between the orbital inclinations of P9 and the TNOs is about $\Delta i=18^{\circ}$. Including the possible uncertainty of the benchmark orbital inclination of the dark P9 $\delta i=5^{\circ}$ \citep{Batygin2}, we take a conservative choice of $\Delta i=25^{\circ}$, which gives $b \sim 8.2$ AU. 

On the other hand, we can also apply the radius of influence $R_{\rm in}$ discussed in \citet{Bate} to characterize the value of the impact parameter (i.e. $b \approx R_{\rm in}$). The radius of influence defines the region where the incoming TNO switches from a two-body problem with central mass $M_{\odot}$ to a two-body problem with central mass $M_9$ in the matched conics approximation \citep{Napier}. Based on this approximation, the impact parameter is given by \citep{Bate}
\begin{equation}
b=R_{\rm in}=a_9\left(\frac{M_9}{M_{\odot}} \right)^{2/5}.
\end{equation}
Using our benchmark parameters, the dark P9 can capture any TNOs moving within the distance of $b \sim 5.3$ AU. To get a more conservative estimation, in the followings, we adopt the value of $b=5.3$ AU as the impact parameter. In view of this, the dark P9 can create a `capturing volume' when it is orbiting about the sun. All of the TNOs inside this capturing volume would be likely captured by the dark P9. The capturing volume is given by
\begin{equation}
V=\left(2 \pi a_9 \sqrt{1-\frac{e_9^2}{2}} \right)(\pi b^2)=2\pi^2 b^2a_9 \sqrt{1-\frac{e_9^2}{2}} \approx 2.5 \times 10^5~{\rm AU^3}.
\end{equation}

Generally speaking, very large TNOs (with size $\ge 500$ km) would be easier for us to identify. Based on the catalog compiled by the International Astronomical Union, there are four TNOs with size $\ge 500$ km (assuming a standard asteroid albedo $p=0.1$) and orbital semi-major axis $a=100-1000$ AU. The number of very large TNOs can provide a standard reference for estimating the amount of TNOs with different sizes inside the scattered disk region. 

Consider the region of the scattered disk for $a=100-1000$ AU. Based on the TNO catalog, all of the reported TNOs with $a \le 1000$ AU are located within a scale disk thickness of 72.5 AU above and below the P9 orbital plane. We therefore consider the volume of the scattered disk $V_d \sim (2\times 72.5)\pi(1000^2-100^2) \approx 4.5 \times 10^8$ AU$^3$. Assuming the distribution of asteroid size is same as that in Kuiper Belt $dN/dD \propto D^{-q}$ \citep{Fraser}. This size distribution in Kuiper Belt is well represented by a broken power law in $D$ for large and small Kuiper Belt objects. For cold Kuiper Belt objects, the slope $q$ for large objects (with size $D \ge 140$ km) is $q=8.2 \pm 1.5$ while $q=2.9 \pm 0.3$ for $D<140$ km \citep{Fraser}. Since there are four TNOs with size $\ge 500$ km, taking $q=8.2$, the average number density of TNOs with size $D \ge 140$ km inside $V_d$ is $8.5 \times 10^{-5}$ AU$^{-3}$.

Since the capturing volume is $2.5 \times 10^5$ AU$^3$, the average number of TNOs with size $D \ge 140$ km captured is about 20. Note that this number is close to the typical number of satellites found in Jovian planets. In fact, the Jovian planets are somewhat close to each other so that the gravitational perturbation effect is significant. This would reduce the capturing volume and the number of satellites. However, there is almost no massive perturber for P9. The closest massive object Sedna (semi-major axis $a\sim 500$ AU) has a relatively small mass $\sim 10^{-3}M_{\oplus}$ only, which cannot affect the capturing volume significantly. Therefore, we expect that there is a considerable amount of captured TNOs to form a satellite system for P9, like the satellite systems in Jovian planets. 

\section{The tidal heating model}
Consider a fiducial radius of the satellite $R=D/2=100$ km. For simplicity, let's assume that the satellite is spherical in shape.  The tidal force on the satellite is large when the satellite is close to P9. The Roche limit is $\sim 2 \times 10^4$ km if we assume the density of the satellite to be $\rho=1$ g/cm$^3$. For Uranus and Neptune, which have mass similar to the dark P9, the range of the orbital semi-major of the satellites is $a_s \sim 5\times 10^4-5\times 10^7$ km. In the followings, we will mainly consider the range of the orbital semi-major axis $a_s=10^5-10^6$ km. Note that captured objects generally have large semi-major axis and eccentricity initially \citep{Goulinski,Napier}. However, orbital evolution through tidal effects would further decrease the values of semi-major axis and eccentricity (see the discussion below).

The equilibrium temperature due to solar luminosity is approximately given by
\begin{equation}
T \approx 54.8 \sqrt{\frac{26}{a_9}}~{\rm K},
\end{equation}
where we have neglected the albedo and the phase integral \citep{Stansberry}. For $a_9=450$ AU, we get $T=13$ K. However, if the satellite is very close to P9, the tidal heating effect would be very significant. The tidal heating model has been discussed for more than 50 years \citep{Goldreich}. In general, the tidal heating rate can be calculated by \citep{Segatz,Lainey,Renaud}
\begin{equation}
\dot{E}=\frac{21C}{2}\frac{(Rn)^5e_s^2}{G},
\end{equation}
where $n=\sqrt{GM_9/a_s^3}$ is the mean orbital motion, and $e_s$ is the eccentricity of the satellite orbit \citep{Segatz}. Here, the constant $C$ is related to the Love number $k_2$ and the quality factor $Q$ which reflects the physical properties (e.g. elastic rigidity) of the satellite \citep{Segatz,Lainey,Hussmann}. However, the value of $C$ for the satellite is uncertain. Theoretical prediction shows that the value of $C$ should be lower than 0.06 for high density satellite core \citep{Kervazo}. We adopt the value revealed from the observational data of the Jupiter's moon Io $C \approx 0.02$ \citep{Lainey}. In equilibrium, the tidal heating rate would be equal to the radiation cooling rate. Therefore, we have
\begin{equation}
T=\left(\frac{\dot{E}}{4\pi \sigma_s \epsilon_{\nu} R^2} \right)^{1/4},
\end{equation}
where $\sigma_s$ is the Stefan-Boltzmann constant and $\epsilon_{\nu}$ is the gray-emissivity. For simplicity, we assume $\epsilon_{\nu}=1$ here. 

In Fig.~1 and Fig.~2, we plot the equilibrium temperature as a function of $a_s$, for different values of $R$ and $e_s$, respectively. We can see that the temperature can be quite high for some values of $a_s$, $R$ and $e_s$. Generally speaking, smaller value of $a_s$ and larger values of $R$ and $e_s$ can give a higher equilibrium temperature. For the fiducial values of $a_s=10^5$ km, $R=100$ km and $e_s=0.5$, we get $\dot{E}=1.4\times 10^{12}$ W. The equilibrium temperature of the satellite is about 119 K, which can emit significant amount of radio radiation with frequency $\nu > 100$ GHz. Besides, we can estimate the time required for the satellite to heat up from 10 K to 100 K. Assuming a typical specific heat capacity for the satellite $c_s=1000$ J kg$^{-1}$ K$^{-1}$, the time required is $\sim 10^4$ yrs for the fiducial parameters used. 

In the followings, we estimate the thermal radio flux emitted by the satellite with the fiducial parameters. The thermal radio flux density is given by
\begin{equation}
S_{\nu}=\int \frac{2h\nu^3}{c^2(e^{h\nu/kT}-1)}d\Omega \approx \frac{2\pi h\nu^3}{c^2(e^{h\nu/kT}-1)} \left(\frac{R}{a_9} \right)^2.
\end{equation}
Therefore, we can get the expected thermal radio flux density as a function of $\nu$ for the fiducial parameters (see Fig.~3). The radio flux density is $\sim 2$ $\mu$Jy for $\nu=300$ GHz. The observable limit for the most sensitive sub-mm interferometer (e.g. Atacama Large Millimeter Array ALMA) is around 1 $\mu$Jy at $\nu=100-300$ GHz. Hence, it is feasible to observe this small flux using current observational technologies. For lower frequencies, the expected radio flux density is $S_{\nu} \approx 10$ nJy at $\nu=20$ GHz. This can be observable by the future SKA radio interferometer.

Moreover, the thermal radio flux density $S_{\nu}$ is proportional to the frequency $\nu^2$. This can be differentiable from the normal background radio flux, which is usually modelled by $S_{\nu} \propto \nu^{-\alpha}$ with $\alpha>0$. In other words, by obtaining the radio spectrum emitted from the region of the dark P9, if we can detect a relatively strong thermal radio spectrum ($S_{\nu} \propto \nu^2$), this would be a solid evidence to verify the P9 hypothesis because there is no other astrophysical mechanism which can increase the temperature of a distant object to more than 50 K. For the conventional P9 model (not a dark object), the expected radio flux emitted by P9 should be $\sim$ mJy at 200 GHz \citep{Naess}, which is 1000 times larger than that of a satellite. In any case, either if we can detect mJy signal from P9 or $\mu$Jy signal from the satellite, the P9 hypothesis can be verified. Besides, if there is any potential signal received from P9 or the satellites, we can track the source for a couple of years to see whether the signal would follow a nearly Keplerian orbit over time or not. This can further provide a smoking-gun evidence to verify the P9 hypothesis.

Previous studies have constrained the possible range of location for P9 \citep{Batygin2,Fienga,Socas}. A recent study has further constrained the exact location of P9 to R.A. $(48.2\pm 4)^{\circ}$ and DEC $(10.3 \pm 1.8)^{\circ}$ \citep{Socas}. Such a small constrained region can make the observation much easier. The telescopes or interferometers used can focus on the target region for a very long exposure time to gain enough sensitivity to detect the potential thermal signals.

Note that the tidal heating rate gained by the satellite originates from the loss rate of the gravitational potential energy of the P9-satellite system. The eccentricity would gradually decrease so that the tidal heating rate would also decrease. The eccentricity fractional change rate is given by
\begin{equation}
\frac{|\dot{e}_s|}{e_s}= \left(\frac{e_s^2-1}{2e_s^2} \right) \frac{\dot{E}}{E}.
\end{equation}
The time scale for the eccentricity shrinking is $\tau \sim |e_s/\dot{e}_s|$, which is about 0.6 Myrs for the fiducial parameters. This timescale is short compared to the age of the solar system. In fact, there is a compromise between having the orbital parameters of the satellites such that the radio emission is detectable (e.g. with small $a_s$) and sufficiently long-lived to make the higher detection probability (e.g. with large $a_s$). Here, the range of $a_s$ we considered ($a_s=10^5-10^6$ km) is almost the optimal for examination. Nevertheless, the relatively short eccentricity shrinking timescale would not be a big problem if the satellite capture event is a recent event. Also, as we have shown that the satellite capture is not a rare event, there would be more than one satellite with size $>140$ km at $a_s \sim 10^5$ km. Therefore, we expect that such a thermal radio signal of the satellite may still be observed. 

\section{Discussion}
In this article, we have demonstrated a theoretical framework to predict the possible observable signal from the P9-satellite system. If the dark P9 has a satellite system, the only current feasible observation is to detect the possible signals from the satellites. We have shown that if a satellite with a typical size $\sim 100$ km with average orbital radius $a_s \sim 10^5$ km from the dark P9, the temperature can be as large as $\sim 100$ K due to tidal heating effect. For such a high temperature, the satellite can emit strong enough thermal radio flux ($\sim 1$ $\mu$Jy at 100-300 GHz) that can be observed by ALMA. Moreover, the specific thermal radio spectrum $S_{\nu} \propto \nu^2$ could be easily differentiable from the background radio flux so that it can provide a smoking-gun evidence for the P9 hypothesis. The only possible reason for the existence of $\sim 100$ K object at $\sim 450$ AU from the sun is that it is a satellite of a host planet. It is because a host dwarf planet or a minor planet does not have enough mass to heat up the satellite to $\sim 100$ K. 

As we have shown above, there are a lot of TNOs with size $>140$ km in the scattered disk region. Therefore, the chance for these large TNOs (with $R \sim 100$ km) captured by P9 is not low. Besides, based on the example of Uranus ($\approx 14M_{\oplus}$), at least 13 satellites are located within $10^5$ km, which suggests that our fiducial value of $a_s=10^5$ km is a reasonable choice of consideration. For the eccentricity, simulations show that most of the captured objects would be orbiting with a very high eccentricity $\approx 1$ \citep{Goulinski}. Therefore, our fiducial value $e_s=0.5$ is a conservative choice of estimation. 

Since no optical and radio signals have been detected so far for P9, the suggestion of P9 being a PBH has become a hot topic recently. There are some suggestions to send detectors to visit the alleged PBH P9 \citep{Witten,Hibberd}. It would be very exciting because this may be our only chance to visit a black hole within our approachable distance. Nevertheless, we need to wait for at least 10 years for the detectors to arrive the PBH P9. Some other studies have proposed to detect P9 by gravitational lensing \citep{Philippov,Schneider,Domenech}. However, the mass of P9 is very small so that it requires a very sensitive measurement for the short-live lensing event, which may not be very easy to get any good confirmation. A recent study has proposed a narrow possible locations of P9 \citep{Socas}. If P9 is a dark object and it has a satellite system, our proposal can directly observe the potential thermal signals emitted by the satellites now. Therefore, this would be a timely and effective method to confirm the P9 hypothesis and verify whether P9 is a dark object or not.

\begin{figure}
\vskip 10mm
\includegraphics[width=140mm]{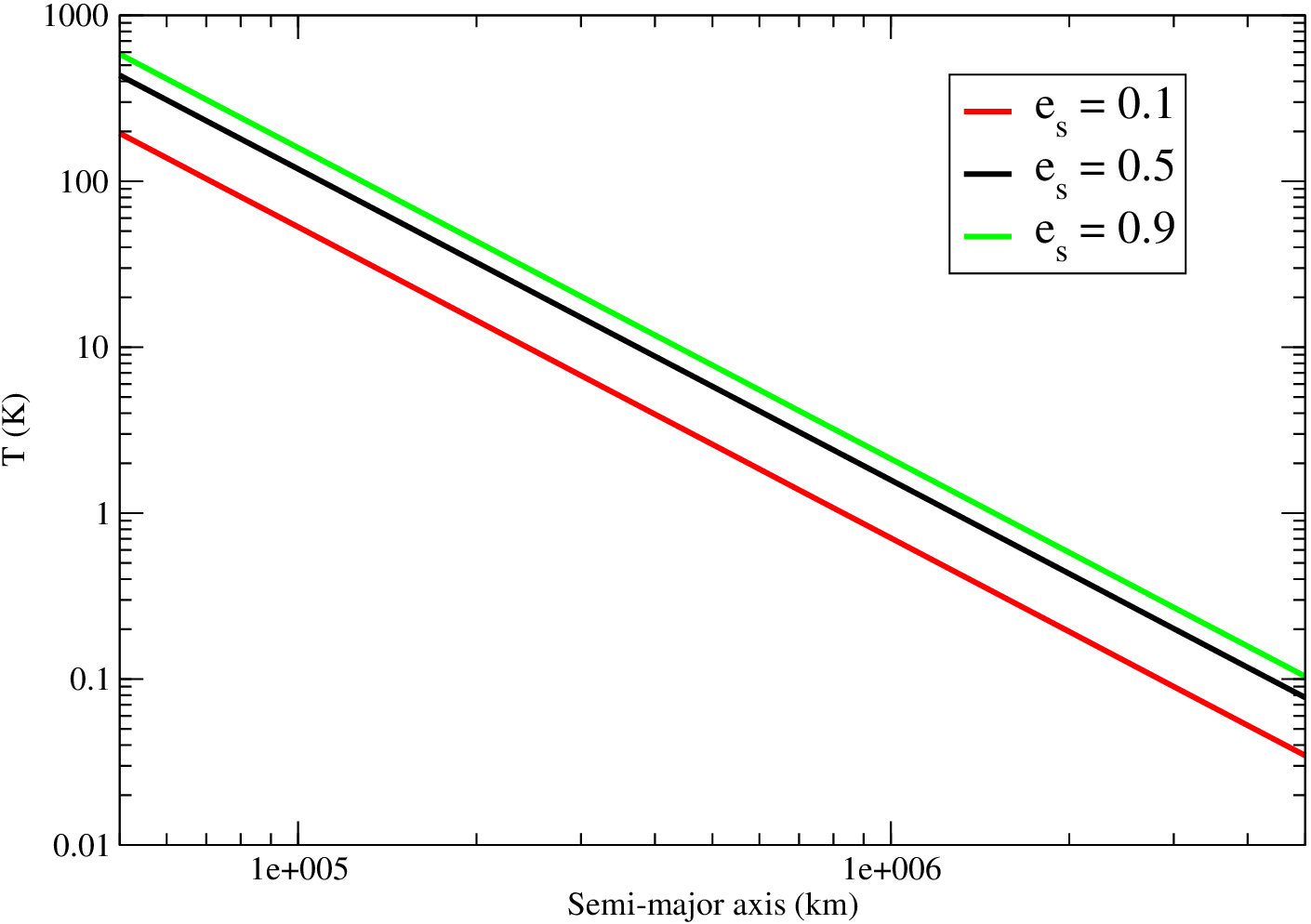}
\caption{The colored lines indicate the predicted temperature $T$ of the satellite for different values of orbital eccentricity ($e_s=0.1$, $e_s=0.5$ and $e_s=0.9$). Here, we have neglected the solar heating effect and we have assumed $R=100$ km.}
\label{Fig1}
\vskip 5mm
\end{figure}

\begin{figure}
\vskip 10mm
\includegraphics[width=140mm]{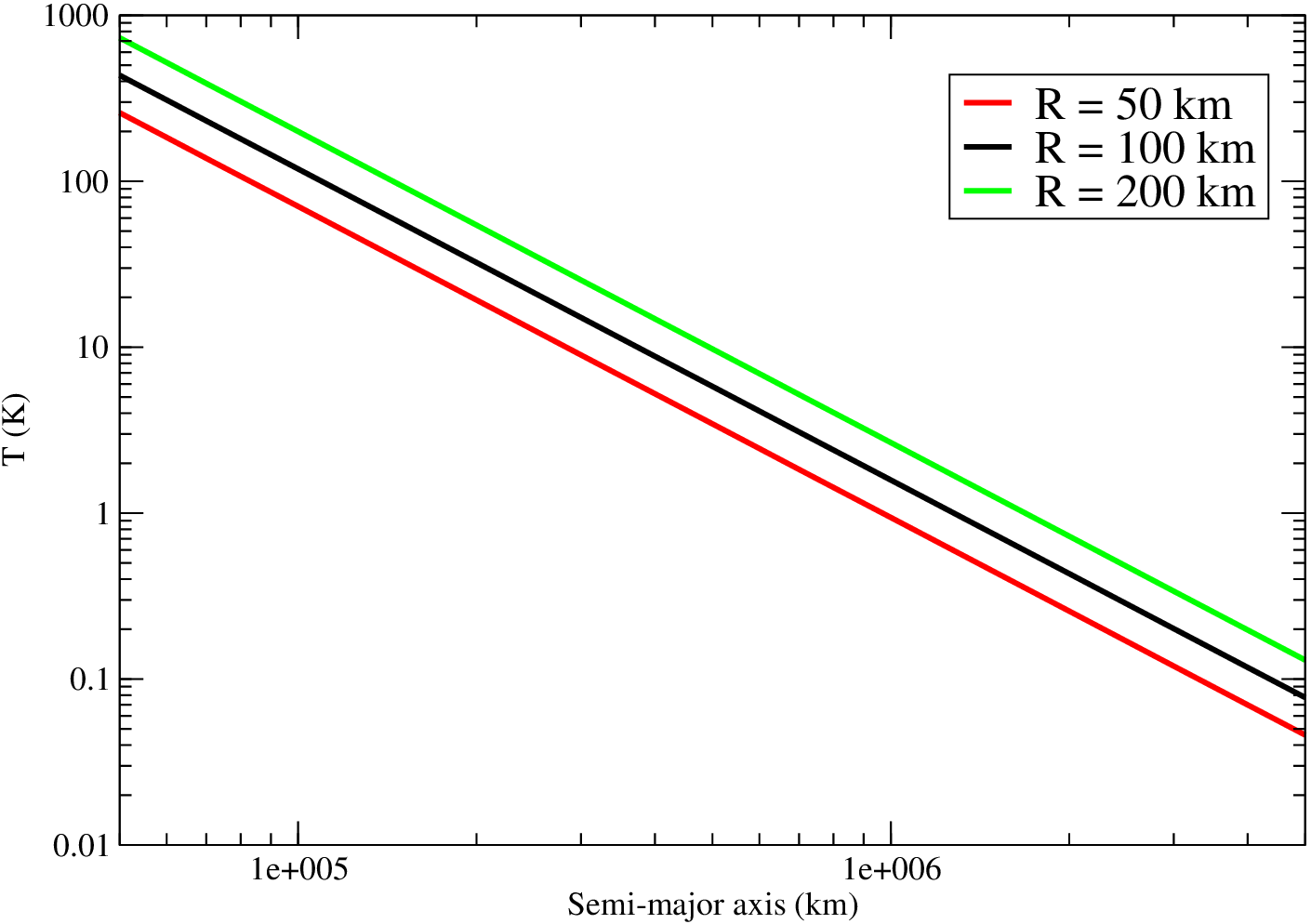}
\caption{The colored lines indicate the predicted temperature $T$ of the satellite for different values of satellite radii ($R=50$ km, $R=100$ km and $R=200$ km). Here, we have neglected the solar heating effect and we have assumed $e_s=0.5$.}
\label{Fig2}
\vskip 5mm
\end{figure}

\begin{figure}
\vskip 10mm
\includegraphics[width=140mm]{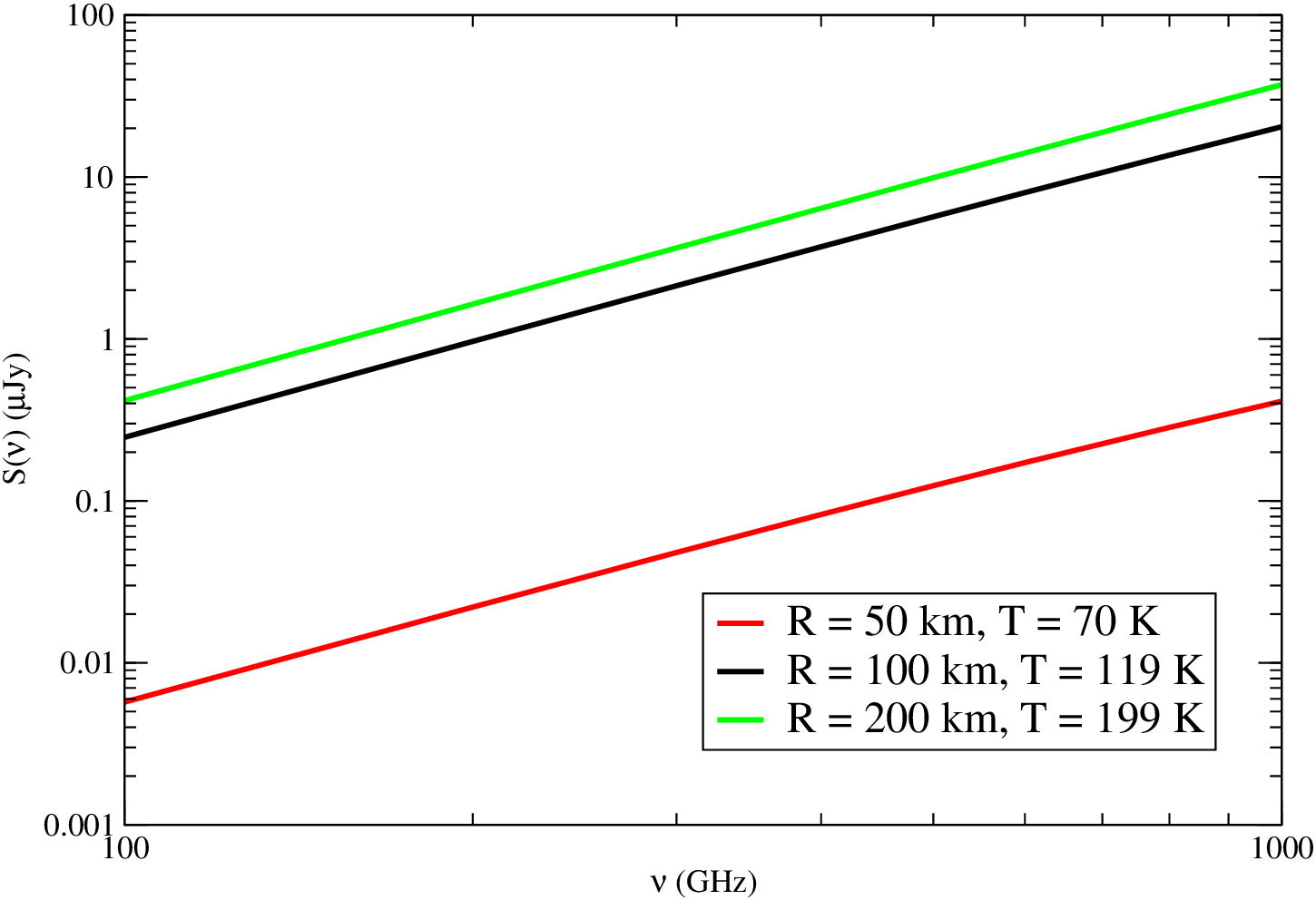}
\caption{The colored lines indicate the predicted thermal radio flux density $S(\nu)$ against $\nu$ for different values of satellite radii ($R=50$ km, $R=100$ km and $R=200$ km). Here, we have assumed $a_s=10^5$ km and $e_s=0.5$.}
\label{Fig3}
\vskip 5mm
\end{figure}

\section{Acknowledgements}
The work described in this paper was partially supported by a grant from the Research Grants Council of the Hong Kong Special Administrative Region, China (Project No. EdUHK 18300922).

\end{document}